\documentclass[preprint,showpacs,preprintnumbers,amsmath,amssymb]{revtex4}

\usepackage{graphicx}
\usepackage{dcolumn}
\usepackage{bm}

\begin{document}

\preprint{APS/123-QED}

\title{Ba(Zn$_{1-2x}$Mn$_x$Co$_x$)$_2$As$_2$: A Bulk Form Diluted Magnetic
Semiconductor with n-type Carriers}

\author{Huiyuan Man$^{1}$, Cui Ding$^{1}$, Shengli Guo$^{1}$, Guoxiang Zhi$^{1}$, Xin Gong$^{1}$, Quan Wang$^{1}$, Hangdong Wang$^{2}$, Bin Chen$^{2}$, F.L.
Ning$^{1,}$}\email{ningfl@zju.edu.cn}

\affiliation{$^{1}$Department of Physics, Zhejiang University,
Hangzhou 310027, China} \affiliation{$^{2}$Department of Physics,
Hangzhou Normal University, Hangzhou 310016, China}

\date{\today}


\begin{abstract}
We report the synthesis and characterization of bulk form diluted
magnetic semiconductors Ba(Zn$_{1-2x}$Mn$_x$Co$_x$)$_2$As$_2$ (0
$\leqslant$ $x$ $\leqslant$ 0.15) with a crystal structure identical
to that of 122-type Fe-based superconductors. Mn and Co co-doping
into the parent compound BaZn$_2$As$_2$ results in a ferromagnetic
ordering below $T_C$ $\sim$ 80 K. Hall effect measurements indicate
that the carrier are n-type with the density of $\sim$
10$^{17}$/cm$^3$. The common crystal structure and excellent lattice
matching between the p-type ferromagnetic
Ba$_{1-y}$K$_y$(Zn$_{1-x}$Mn$_x$)$_2$As$_2$, the n-type
ferromagnetic Ba(Zn$_{1-2x}$Mn$_x$Co$_x$)$_2$As$_2$, the
antiferrmagnetic BaMn$_2$As$_2$ and the superconducting
Ba(Fe$_{1-x}$Co$_x$)$_2$As$_2$ systems make it possible to make
various junctions between these systems through the As layer.
\end{abstract}


\maketitle


The research into diluted magnetic semiconductors (DMS) has been
explosive since the observation of ferromagnetic ordering in III-V
(Ga,Mn)As thin-film by Ohno eta al
\cite{Ohno}\cite{Jungwirth,Dietl1,Zutic}. After almost 20 years of
efforts, the highest Curie temperature, $T_C$, has been reported as
$\sim$200 K with Mn doping levels of $\sim$12 $\%$ in (Ga,Mn)As
\cite{ZhaoJH1,ZhaoJH2}. It has been proposed theoretically that the
application of spintronics will become possible once $T_C$ reaches
room temperature \cite{Zutic}. Improving $T_C$ is, however, hindered
by an inherent difficulty: the mismatch of valences of Ga$^{3+}$ and
Mn$^{2+}$ prohibits the fabrication of bulk form specimens and
homogenous thin films with higher Mn doping levels. On the other
hand, Mn substitution for Ga not only acts as a local moment, but
also donates a hole, i.e., p-type carriers. Furthermore, some Mn
impurities enter interstitial sites, which makes it difficult to
determine precisely the amount of Mn that substitutes Ga
\cite{Jungwirth}. The absence of decisive determination of the
actual amount of Mn that substituted Ga also makes it difficult to
measure where holes reside in: the impurity band or the valence band
\cite{Dobrowolska}.

Prior to the research of (Ga,Mn)As, a large group of II-VI family of
DMS have also been extensively studies. Because Mn ions have the
same valence as that of Zn ions, the chemical solubility can be very
high, i.e., $\sim$70\% in (Zn$_{1-x}$Mn$_x$)Se
\cite{Pajaczkowska,Furdyna}, and bulk form specimens are available.
However, it is difficult to control the carrier density, which is as
low as 10$^{17}$ cm$^{-3}$, and the type of carriers
\cite{Wojtowicz,Morkoc}. The magnetic moment size is as small as
0.01 $\mu_B$/Mn in most II-VI DMS\cite{Furdyna,Shand}. In general,
the ground state is believed to be a spin-glass, instead of a long
range ferromagnetic ordering.

An pioneer theoretical work proposed by Masek et al has shown that
the I-II-V direct-gap semiconductor LiZnAs is a good candidate for
fabrication of next generation of DMS \cite{Masek}. LiZnAs is a
direct gap semiconductor with a band gap of 1.6 eV
\cite{Bacewicz,Kuriyama1, Kuriyama2,Wei}. It has a cubic structure,
similar to that of zinc-blende GaAs and ZnSe. More interestingly, if
we view the combination of (Li$^{1+}$Zn$^{2+}$) as Ga$^{3+}$,
(LiZn)As becomes GaAs; alternatively, if we view the combination of
(Li$^{1+}$As$^{3-}$) as Se$^{2-}$, Zn(LiAs) becomes ZnSe. From the
view of synthesis: The I-II-V semiconductor LiZnAs has two superior
advantages over III-V and II-VI semiconductors: (1) the isovalent
substitution of Mn for Zn overcomes the small chemical solubility
encountered in (Ga,Mn)As; (2) the carrier type and density can be
controlled by off-stoichiometry of Li concentrations., i.e., extra
Li introduce electrons and Li deficiency bring holes.

Recently, Deng et al. successfully synthesized two bulk I-II-V DMS
systems, Li(Zn,Mn)As \cite{Deng1} and Li(Zn,Mn)P \cite{Deng2}, with
$T_C$ $\sim$ 50 K. It has been shown by muon spin relaxation
($\mu$SR) and nuclear magnetic resonance (NMR) that Mn atoms are
homogenously doped into the compound, and the ferromagnetic ordering
is truly arising from the Mn atom that substituted into the ionic Zn
sites, not from magnetic impurities or clusters. The I-II-V DMSs
have advantages of decoupling spins and carriers, where spins are
introduced by Mn atoms and carriers are created by off-stoichiometry
of Li concentrations. This advantage makes it possible to precisely
control the amount of spins and carriers, and investigate their
individual effects on the ferromagnetic ordering. The puzzle in
Li(Zn,Mn)As and Li(Zn,Mn)P DMS, however, is that the carrier is
always p-type even excess Li are introduced during the synthesis
process. This is in contrast with the intuitive expectation that
excess Li will render a $n$-type carriers. There are no convincing
experiments to clarify this issue so far. Based on first-principles
calculations, Deng et al \cite{Deng2} shows that excess Li$^{1+}$
ions are thermodynamically favored to occupy the Zn$^{2+}$ sites,
and each Li$^{1+}$ substitution for Zn$^{2+}$ will introduce a hole
carrier.

Very recently, several more bulk DMS systems with decoupling spins
and charges have been reported. Firstly, Ding et al \cite{Ding1} and
Han et al \cite{HanW} reported the ferromagnetic ordering below
$T_C$ $\sim$ 40 K in ``1111" type (La,Ba)(Zn,Mn)AsO and
(La,Ca)(Zn,Mn)SbO systems; and Yang et al reported the fabrication
of another "1111" type DMS (La,Sr)(Cu,Mn)SO with $T_C$ $\sim$ 210 K
\cite{Yangxj}. Secondly, Zhao et al. reported the ``122" type DMS
systems, (Ba,K)(Zn,Mn)$_{2}$As$_{2}$, which has $T_C$ as high as 180
K \cite{ZhaoK}, and Yang et al observed the ferromagnetic transition
below $T_C$ $\sim$ 17 K and a large negative magnetoresistance in
(Ba,K)(Cd,Mn)$_{2}$As$_{2}$ \cite{Yang2}. It is worth to note that
the Curie temperature of (La,Sr)(Cu,Mn)SO and
(Ba,K)(Zn,Mn)$_{2}$As$_{2}$ polycrystals is already comparable to
the record $T_C$ of (Ga,Mn)As thin films \cite{ZhaoJH1, ZhaoJH2}.
Interestingly, the type of carriers in above bulk form DMSs are all
p-type, and no n-type DMS has been achieved.

\begin{figure}[!htpb] \centering \vspace*{+0.5cm}
\centering
\includegraphics[width=3.3in]{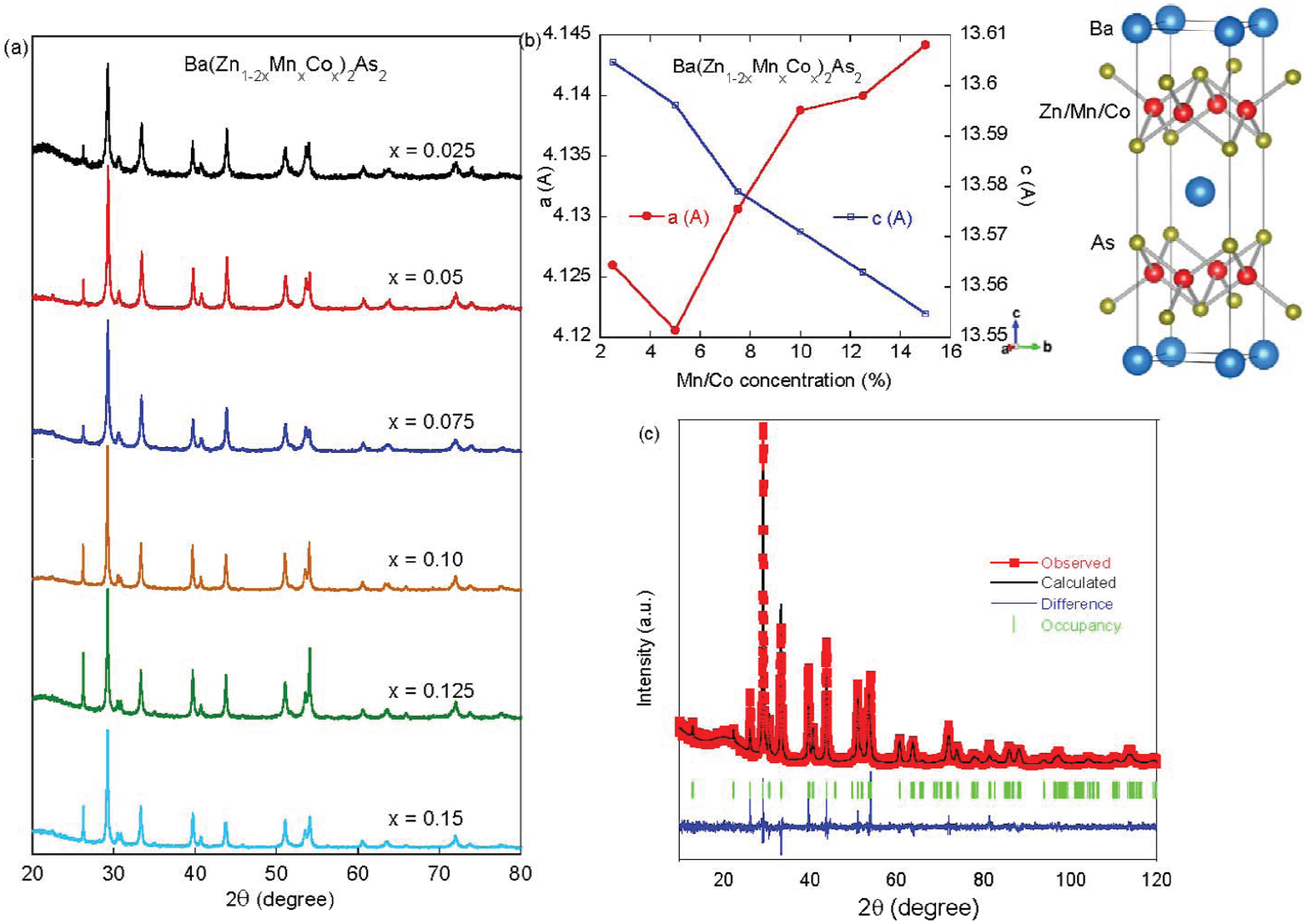}\vspace*{+0.5cm}
\caption{\label{Fig1:epsart} (Color online). (a) The X-ray diffraction patterns of Ba(Zn$_{1-2x}$Mn$_x$Co$_x$)$_2$As$_2$ (0 $\leqslant$ $x$ $\leqslant$ 0.15). (b) The systematic change of the lattice constants a and c. The crystal structure is shown, which is isostructural to 122 Fe-based supercondutors. (c) The XRD diffraction peaks can be well indexed into the tetragonal structure of space group I4/mmm.}
\end{figure}

In this letter, we report the synthesis and characterization of Mn
and Co co-doped Ba(Zn$_{1-2x}$Mn$_x$Co$_x$)$_2$As$_2$ (0 $\leqslant$
$x$ $\leqslant$ 0.15) semiconductors. We found that Mn and Co
co-doping into the parent compound BaZn$_2$As$_2$ results in a
ferromagnetic ordering below $T_C$ $\sim$ 80 K. The coercive field
is in the order of $\sim$200 Oe, which is comparable to that of the
cubic (Ga,Mn)As, Li(Zn,Mn)As and Li(Zn,Mn)P. The electrical
transport measurements show typical semiconducting behavior. Hall
effect measurements indicate that the carrier are n-type with the
density of $\sim$ 10$^{17}$/cm$^3$. We found that doping Co alone
results in some type of magnetic ordering with a much smaller
saturation moment size. Only co-doped with Mn atoms, a ferromagnetic
ordering with saturation moment size of $\sim$ 0.1 $\mu$B/TM (TM =
Mn, Co) is observed.

We synthesized the polycrystalline specimens
Ba(Zn$_{1-2x}$Mn$_x$Co$_x$)$_2$As$_2$ ($x$ = 0.025, 0.05, 0.075, 0.10, 0.125, 0.15)
by the solid state reaction method. Zn (99.9\%), Mn (99.99\%), Co (99.99\%) and As (99\%) were sintered at 800$^{\circ}$ for 10 hours to make the precursors ZnAs, MnAs and CoAs. Then the mixture of Ba (99.9\%), ZnAs, MnAs and CoAs were slowly heated to 1150$^{\circ}$C in evacuated silica tubes, and held for 60 hours before cooling down to room temperature at the rate of 50$^{\circ}$C/h. The polycrystals were characterized by X-ray diffraction at room temperature and dc magnetization by Quantum Design SQUID. The electrical resistance was measured on sintered pellets with typical four-probe method.

\begin{figure}[!htpb] \centering \vspace*{+0.2cm}
\centering
\includegraphics[width=3.3in]{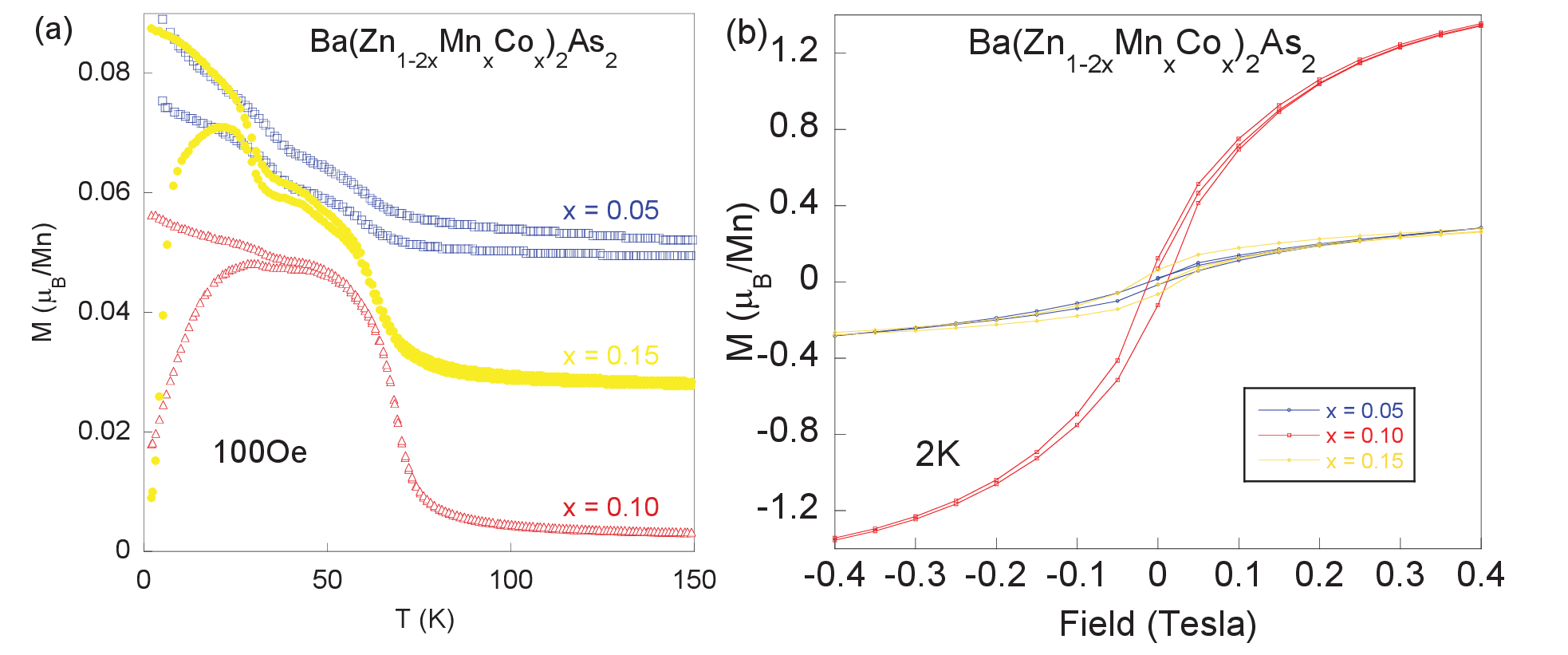}\vspace*{+0.1cm}
\caption{\label{Fig2:epsart} (Color online) (a) The temperature dependent magnetization M for Ba(Zn$_{1-2x}$Mn$_x$Co$_x$)$_2$As$_2$ with x = 0.05, 0.10, 0.15 obtained in the zero field cooling (ZFC) and field cooling (FC) under the external field of 100 Oe. (b) The isothermal magnetization measured at 2 K shows hysteresis loops for this material. The coercive filed increases with increasing doping level.}
\end{figure}

We show the crystal structure of
Ba(Zn$_{1-2x}$Mn$_x$Co$_x$)$_2$As$_2$ and the X-ray diffraction
patterns in Fig. 1. Bragg peaks from the
Ba(Zn$_{0.9}$Mn$_{0.05}$Co$_{0.05}$)$_2$As$_2$ can be well indexed
by a 122 tetragonal structure, isostructural to the 122 type of
Fe-based superconductor Ba(Fe$_{1-x}$Co$_x$)$_2$As$_2$, as shown in
Fig. 1(c). The single phase is conserved with the doping level up to
$x$ = 0.10. Small traces of non-magnetic Zn$_3$As$_2$ impurities
appear for higher level of doping. The lattice constant $a$
monotonically increases from $a$ = 4.120 {\AA} for $x$ = 0.05 to
4.145 {\AA} for $x$ = 0.15, and $c$ monotonically decrease from $c$
= 13.605 {\AA} to 13.554 {\AA}, indicating the successful solid
solution of Cr for Zn.

\begin{figure}[!htpb] \centering \vspace*{-0.1cm}
\centering
\includegraphics[width=3.3in]{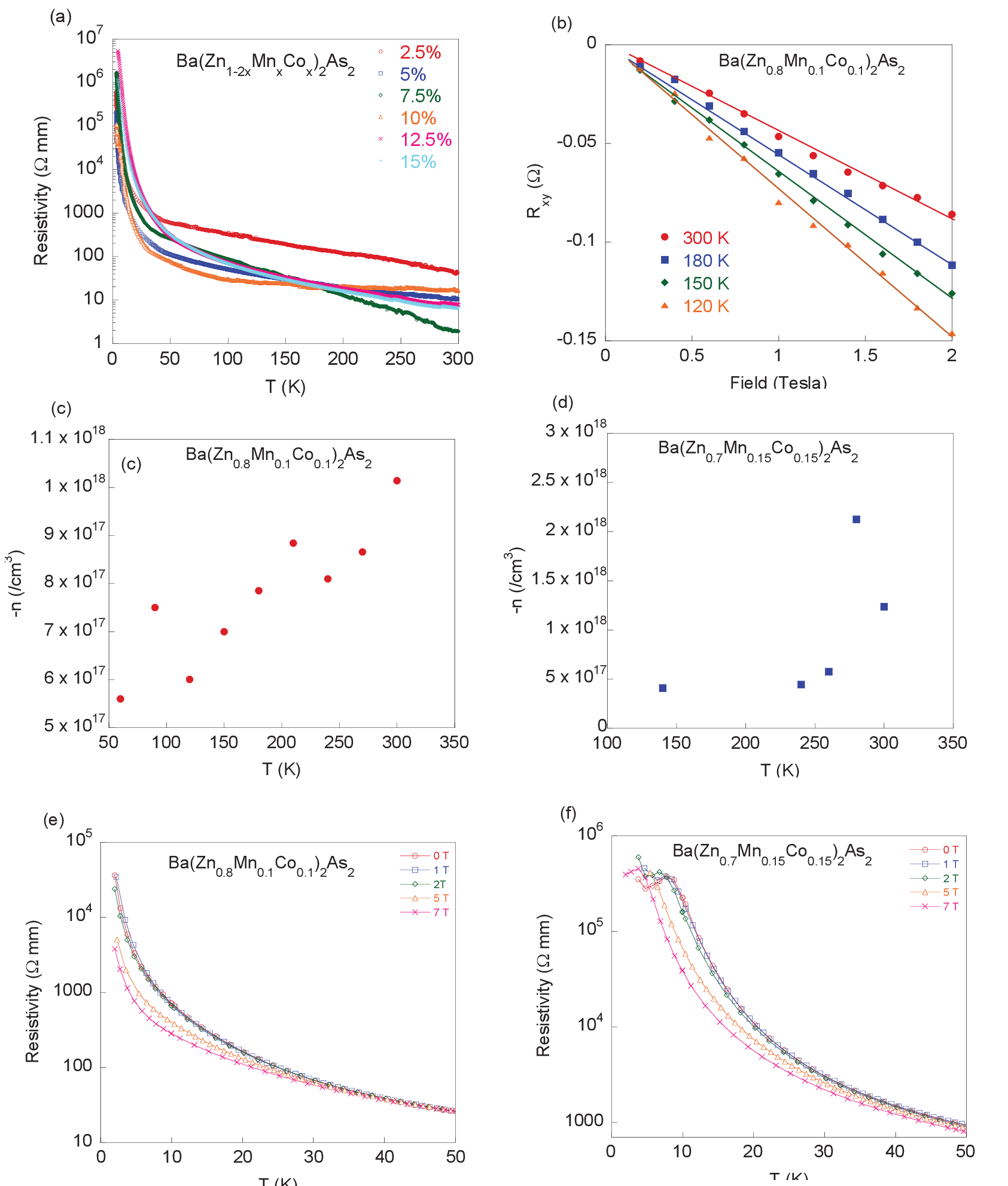}\vspace*{-0.1cm} \\
\caption{\label{Fig3:epsart} (Color online) (a) The temperature dependence of resistivity for Ba(Zn$_{1-2x}$Mn$_x$Co$_x$)$_2$As$_2$. (b) Hall resistance of Ba(Zn$_{0.8}$Mn$_{0.1}$Co$_{0.1}$)$_2$As$_2$ at 300 K, 180 K, 150 K and 120 K with linear fitting lines. T-dependent carrier density of Ba(Zn$_{0.8}$Mn$_{0.1}$Co$_{0.1}$)$_2$As$_2$ (c) and  Ba(Zn$_0.7$Mn$_{0.15}$Co$_{0.15}$)$_2$As$_2$ (d) decreases toward low temperature. The resistivity of Ba(Zn$_{0.8}$Mn$_{0.1}$Co$_{0.1}$)$_2$As$_2$ (e) and  Ba(Zn$_0.7$Mn$_{0.15}$Co$_{0.15}$)$_2$As$_2$ (f) under different external fileds shows a large magnetic resistance.}
\end{figure}

In Fig. 2(a), we show the zero-field cooled (ZFC) and field cooled
(FC) measurements of the $dc$-magnetization $M$ of
Ba(Zn$_{1-2x}$Mn$_x$Co$_x$)$_2$As$_2$ ($x$ = 0.05, 0.10, 0.15) for
$B_{ext}$ = 100 Oe. For the doping of $x$ = 0.05, we observe a
strong increase of M at $T_C$ = 60 K, but the splitting between ZFC
and FC curves is small. With the doping level increasing to $x$ =
0.10, a significant increase in $M$ is observed at the temperature
of $\sim$ 75 K, and the ZFC and FC curves split, indicating that
ferromagnetic ordering is taking place. $T_C$ decreases to $\sim$ 70
K with the doping level of $x$ = 0.15. A close inspection of M
indicates that for $x$ = 0.05 and 0.10, a step has been observed at
$T$ $\sim$ 30 K. This is possibly arising from either a spin state
change of Co 3d electrons, or the inhomogeneous distribution of Mn
or Co atoms in the specimens. We are still working to optimize the
synthesis condition and to improve the sample homogeneity. None the
less, we believe that either reason does not affect the fact that
$T_C$ reaches $\sim$ 80 K at the average doping level of 10 \% Mn
and Co. We fit the temperature dependence of $M$ above $T_C$ to a
Curie-Weiss law. The effective paramagnetic moment is determined to
be 2 $\sim$ 3$\mu _B$/TM (TM = Mn, Co). Currently we do not firmly
know if Co introduces spins during the substitution process, and it
is difficult to determine the spin state of both Mn and Co. In the
p-type DMS mentioned above, Mn ions are of ``+2" valence and its
high spin state have been demonstrated.

In Fig. 2(b), we show the isothermal magnetization of
Ba(Zn$_{1-2x}$Mn$_x$Co$_x$)$_2$As$_2$ ($x$ = 0.05, 0.10, 0.15). For
all doping levels, a parallelogram-shaped hysteresis loop with a
small coercive field has been observed. For $x$ = 0.05, the coercive
field at 2 K is 100 Oe. The coercive field continuously increases to
258 Oe with the doping level increasing to 15\%. The magnitude of
coercive field is comparable to $\sim$ 50 - 100 Oe of the cubic
structural Li$_{1.1}$(Zn$_{0.97}$Mn$_{0.03}$)As \cite{Deng1},
Li$_{1.1}$(Zn$_{0.97}$Mn$_{0.03}$)P \cite{Deng2} and
(Ga$_{0.965}$Mn$_{0.035}$)As \cite{Ohno}, but much smaller than the
$p$ -type Ba$_{1-y}$K$_y$(Zn$_{1-x}$Mn$_x$)$_2$As$_2$ \cite{ZhaoK},
which has an identical two dimensional crystal structure.

In Fig. 3(a), we show the electrical resistivity measured for
Ba(Zn$_{1-2x}$Mn$_x$Co$_x$)$_2$As$_2$ ($x$ = 0.025, 0.05, 0.075,
0.10, 0.125, 0.15). For the parent semiconductor BaZn$_2$As$_2$, the
electrical resistivity displays typical semiconducting behavior.
Small amount of K doping into Ba sites change it into a metal
\cite{ZhaoK}. In the case of Mn and Co co-doped samples, we found
that the resistivity monotonically increases with the decreasing of
temperature for all doping levels. The same type of resistive
behavior has also been observed in (Ga$_{1-x}$Mn$_{x}$)As for $x$
$\leq$ 0.03. It has been ascribed to the scattering of carriers by
magnetic fluctuations through exchange interactions in
(Ga$_{1-x}$Mn$_{x}$)As \cite{Jungwirth}. However, we do not observe
insulator to metal transition when the doping level reaches as high
as 15\%.

We have also conducted the Hall effect measurements for the sample
of Ba(Zn$_{1-2x}$Mn$_x$Co$_x$)$_2$As$_2$ ($x$ = 0.10 and 0.15), and
show the data in Fig. 3(b)-(d). Our results indicate that the
carriers are $n$-type, with an electron density in the order of $n
\sim 5 \times 10^{17}$ cm$^{-3}$. This carrier density is comparable
to that of Li$_{1.1}$Zn$_{1-x}$Mn$_{x}$P \cite{Deng2} and II-VI
DMSs. In the research of Fe-based high temperature superconductors,
Co substitution for Fe in BaFe$_2$As$_2$ introduce electrons, and
superconductivity with $T_c$ $\sim$ 25 K has been achieved. It has
been shown by ARPES (angle-resolved photoemission spectroscopy) that
sufficient amount of Co doping kills the hole Fermi surfaces \cite{Stewart} 
residing in the center of the Brillion zone, and expand the electron
Fermi surfaces \cite{Stewart}. The n-type carriers in
Ba(Zn$_{1-2x}$Mn$_x$Co$_x$)$_2$As$_2$ should be attributed to the Co
substitution for Zn. We also measured the magneto-resistance (MR)
for the doping level of $x$ = 0.10 and 0.15. Under the magnetic
field of 7 Tesla, the magnitude of electrical resistivity at 2 K
decrease -89\% and  -78\% for $x$ = 0.10 and 0.15 samples,
respectively. The negative magnetoresistance can be attributed to
the reduction of disorders among the local spins, and the followed
suppression of magnetic scattering by external field.

In diluted magnetic systems, the magnetic impurities can easily give
rise to spurious features of ``ferromagnetism", such as the
bifurcation of ZFC and FC curves and hysteresis loops \cite{Samarth,
Chambers}. In the Ba(Zn$_{1-2x}$Mn$_x$Co$_x$)$_2$As$_2$ system, the
ordering temperature $T_C$ systematically changes with the variation
of Mn and Co doping levels, indicating that the magnetic ordering is
truly arising from the Mn and Co atoms that substituted for the Zn
atoms in ionic sites. We would observe a single transition
temperature for all doping levels if the magnetic ordering arises
from the same type of magnetic impurity source. $\mu$SR technique is
a powerful tool to determine the magnetic ordered volume at a
microscopic level, as has been demonstrated in the bulk form p-type
DMS materials \cite{Deng1, ZhaoK, Ding1}. We have also conducted
$\mu$SR measurement of the Ba(Zn$_{1-2x}$Mn$_x$Co$_x$)$_2$As$_2$,
and our preliminary $\mu$SR results on 15\% doped specimen indicate
that static ferromagnetic ordering do develops at low temperature.

On the other hand, the bifurcation of ZFC and FC curves and the
hysteresis loops can be found not only in regular ferromagnets
\cite{Ashcroft} but also in spin glasses \cite{Fischer}. The
saturation moment reach $\sim$ 1 $\mu_B$/TM (TM = Mn, Co) in
Ba(Zn$_{1-2x}$Mn$_x$Co$_x$)$_2$As$_2$ under the field of 0.4 Tesla.
The ground state is not likely a spin glass state because the
saturation moment of a spin glasses is usually in the order of 0.01
$\mu_B$/Mn or even smaller. Experimental, neutron scattering can
resolve spatial spin correlations, and is a decisive technique to
distinguish the two cases. We have conducted neutron diffraction
experiment on (Ba$_{0.7}$K$_{0.3}$)(Zn$_{0.9}$Mn$_{0.1}$)$_2$As$_2$
\cite{ZhaoK} and (La$_{0.9}$Sr$_{0.1}$)(Zn$_{0.9}$Mn$_{0.1}$)AsO
\cite{Lu} polycrystalline DMS specimens which have a much larger
saturation moment size, $\sim$ 0.4 - 1 $\mu_B$/Mn. Unfortunately, it
is still difficult to decouple the magnetic and structural Bragg
peaks even at 6 K due to the spatially dilute Mn moments
\cite{Ning}. For the n-type Ba(Zn$_{1-2x}$Mn$_x$Co$_x$)$_2$As$_2$
system, it will be easier to grow single crystals since Mn and Co
are more easily to alloy with Zn. Upon the single crystals are
available, high resolution neutron scattering experiments will be
conducted.

Several theoretical models has been proposed to explain the
ferromagnetism in various diluted magnetic semiconductors and
oxides, such as Zener's model \cite{Dietl2}, percolation of bound
magnetic polarons (BMPs) \cite{Bhatt, Sarma, Coey}, and $d-d$ double
exchange due to hopping between transition metal $d$ states
\cite{Millis}. In the case of Ba(Zn$_{1-2x}$Mn$_x$Co$_x$)$_2$As$_2$,
the resistivity is very large ($\sim$ 10$^5$ $\Omega$ mm) and the
carrier density is as low as $\sim$ 10$^{17}$ cm$^{-3}$. The origin
of the ferromagnetism seems more amenable with the BMPs model.

In summary, we reported the synthesis and characterization of bulk
form diluted magnetic semiconductors
Ba(Zn$_{1-2x}$Mn$_x$Co$_x$)$_2$As$_2$ with the ordering temperature
as high as $\sim$ 80 K. It is the first time that Mn and Co atoms
are co-doped into the BaZn$_2$As$_2$ semiconductor and a
ferromagnetic ordering is observed. The Hall effect measurements
indicate that the carrier is $n$-type with a density in the order of
10$^{17}$/cm$^{-3}$. The magneto-resistance reaches to $\sim$ -89\%
under the external field of 7 Tesla. Finally, we would like to
emphasize that the common crystal structure and excellent lattice
matching between the p-type ferromagnetic
Ba$_{1-y}$K$_y$(Zn$_{1-x}$Mn$_x$)$_2$As$_2$, the n-type
ferromagnetic Ba(Zn$_{1-2x}$Mn$_x$Co$_x$)$_2$As$_2$, the
antiferrmagnetic BaMn$_2$As$_2$ and the superconducting
Ba(Fe$_{1-x}$Co$_x$)$_2$As$_2$ systems make it possible to make
various junctions between these systems through the As layer.

The work at Zhejiang University was supported by National Basic
Research Program of China (No. 2011CBA00103, No. 2014CB921203), NSFC
(No. 11274268); F.L. Ning acknowledges helpful discussion with C.Q.
Jin and Y.J. Uemura.
\\



\end{document}